\documentclass[prl,twocolumn,superscriptaddress]{revtex4-1}

\usepackage{color}
\usepackage[english]{babel}
\usepackage[T1]{fontenc}
\usepackage[latin9]{inputenc}
\usepackage{latexsym}
\usepackage{float}
\usepackage{amsmath}
\usepackage{graphicx}
\usepackage{times}   
\usepackage{esint}
\usepackage[unicode=true,pdfusetitle,
 bookmarks=false,colorlinks=true,citecolor=blue,urlcolor=blue,linkcolor=red]{hyperref}

\makeatletter

\@ifundefined{textcolor}{}
{%
 \definecolor{BLACK}{gray}{0}
 \definecolor{WHITE}{gray}{1}
 \definecolor{RED}{rgb}{1,0,0}
 \definecolor{GREEN}{rgb}{0,1,0}
 \definecolor{BLUE}{rgb}{0,0,1}
 \definecolor{CYAN}{cmyk}{1,0,0,0}
 \definecolor{MAGENTA}{cmyk}{0,1,0,0}
 \definecolor{YELLOW}{cmyk}{0,0,1,0}
}

\@ifundefined{date}{}{\date{}}
\AtBeginDocument{
  
}
\makeatother

\newcommand{\SAVE}[1]{}

\newcommand{\prlsec}[1]{\emph{#1---}}

\newcommand{\Scal}{{\mathcal S}}

\begin{document}
\renewcommand\abstractname{}

\title{Quantum versus classical effects at zero and finite temperature in the quantum pyrochlore Yb$_2$Ti$_2$O$_7$} 
\author{Hitesh J. Changlani}
\affiliation{Department of Physics and Astronomy, Johns Hopkins University, Baltimore, MD 21218 
and Institute for Quantum Matter, Johns Hopkins University, Baltimore, MD 21218}
\date{\today}

\begin{abstract}
We study the finite temperature properties of the candidate quantum spin ice material Yb$_2$Ti$_2$O$_7$ 
within the framework of an anisotropic nearest-neighbor spin $1/2$ model on the pyrochlore lattice. 
Using a combination of finite temperature Lanczos and classical Monte Carlo 
methods, we highlight the importance of quantum mechanical effects
for establishing the existence and location of the low-temperature 
ordering transition. We perform simulations of the $32$ site cluster, 
which capture the essential features of the specific heat curve seen in the cleanest known samples of this material.  
Focusing on recent experimental findings [A. Scheie et al., Phys. Rev. Lett. 119, 127201 (2017) 
and J. D. Thompson et al., Phys. Rev. Lett. 119, 057203 (2017)], we then address the question of 
how the phase boundary between the ferromagnetic and paramagnetic phases changes when subjected to a magnetic field. 
We find that the quantum calculations explain discrepancies observed with a completely classical 
treatment and show that Yb$_2$Ti$_2$O$_7$ displays significant renormalization effects, 
which are at the heart of its reentrant lobed phase diagram. Finally, we develop a qualitative understanding 
of the existence of a ferromagnet by relating it to its counterpart that exists in the vicinity of the classical ice manifold.
\end{abstract}

\maketitle
\prlsec{Introduction} Frustrated magnets constitute a fertile hunting ground for discovering 
unconventional states of matter, including spin liquids with topological properties. 
The presence of multiple competing energy scales is at the heart of several contentious issues - 
ranging from the precise knowledge of the low-energy effective Hamiltonian to the reliable 
determination of the low-energy properties. While spin liquids are desirable, 
"order-by-disorder" effects~\cite{Henley_order_by_disorder, Villain} typically 
dominate leading to magnetically ordered or valence bond states. 
However, combined experimental and theoretical efforts determined Hamiltonian parameters for Yb$_2$Ti$_2$O$_7$~(YbTO)~\cite{Ross_PRX, Savary_gaugeMFT, Bloete}, and suggested a quantum spin ice (spin liquid) ground state~\cite{Hermele_Fisher_Balents}, 
possibly circumventing the issues above. 
This phase is qualitatively described as a quantum superposition of configurations in which the spins are constrained to point into or out of 
tetrahedra of the pyrochlore lattice, with a two-in-two-out "ice rule"~\cite{Bernal_Fowler, Ramirez1999, Gingras_review}, 
a schematic of one such configuration is depicted in Fig.~\ref{fig:32site}. 
Defects (spin flips) in this rule produce a pair of magnetic "monopoles"; the analogy with electrodynamics led to the theoretical 
prediction of exotic gapless excitations or "photons"~\cite{Hermele_Fisher_Balents}.
This fuelled several other studies~\cite{Pan_Armitage, Wan_Tchernyshyov, Applegate, Robert, Jaubert_multiphase} to understand the true nature of YbTO. 
 
Real materials always have some disorder; thus it is important to 
clarify the nature of the ground state theoretically.
Multiple works have addressed the issue at the classical level~\cite{Shannon_edge, Jaubert_multiphase,Robert,Scheie2017}, 
but quantum treatments have been limited to mean field theories~\cite{Chang_Onoda}, 
small clusters~\cite{Onoda_quantum, Jaubert_multiphase}, high temperature approaches~\cite{Hayre,Applegate, Jaubert_multiphase} or 
sign problem free parameter sets~\cite{Shannon_QMC,Onoda_QMC}; the latter may not represent YbTO. 

\begin{figure}[htpb]
\centering
\includegraphics[width=0.85\linewidth]{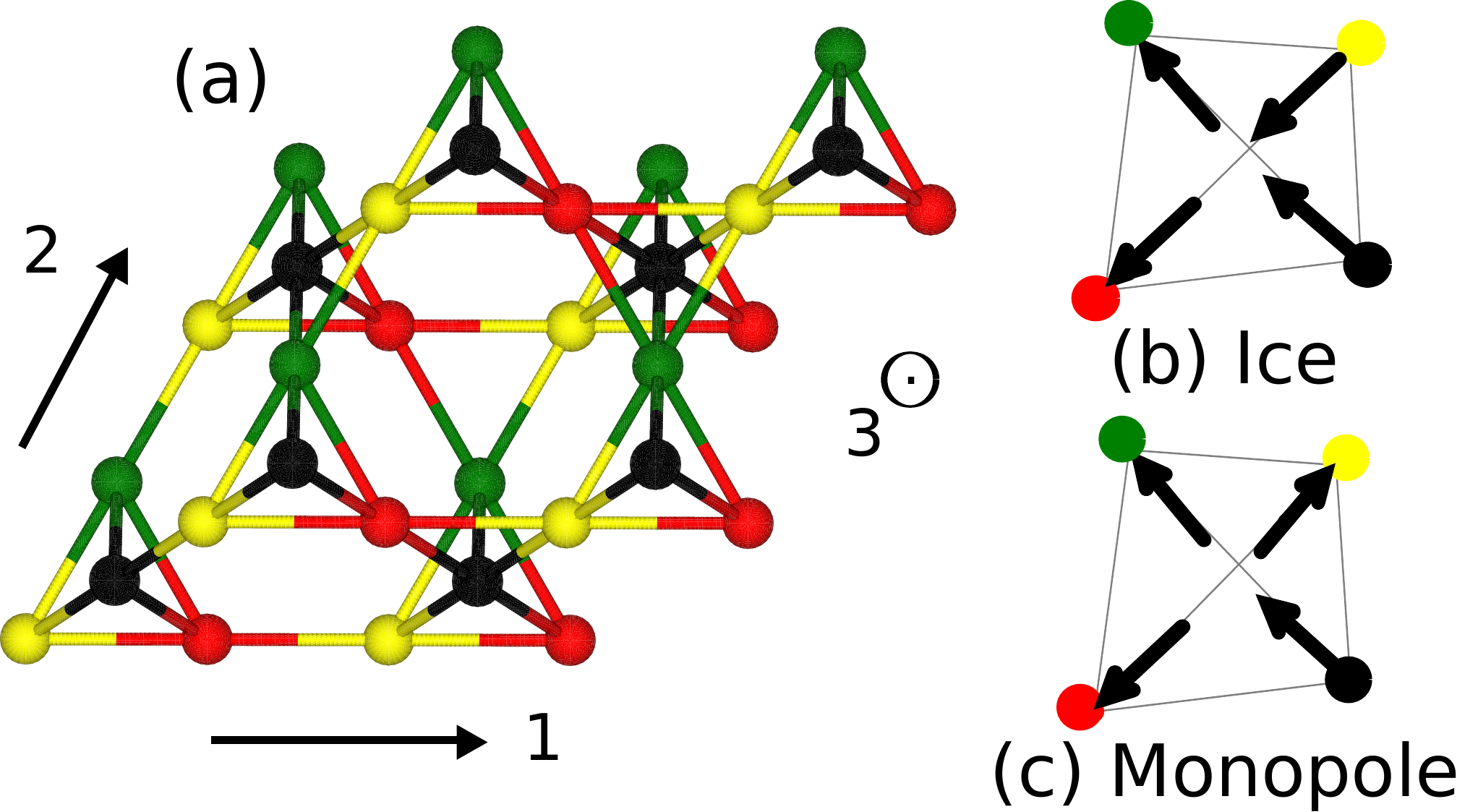}
\caption{(a) $32$ site unit cell of the pyrochlore lattice used for the calculations in the paper, 
as viewed along the global [111] direction 
with periodic boundary conditions along the $1$,$2$ and $3$ directions. The four colors represent 
the four FCC sublattices. (b) and (c) show 
representative configurations of four spins on a single tetrahedron
(b) satisfying the two-in two-out ice rule and (c) a monopole associated with a defect in the local ice rule.}
\label{fig:32site} 
\end{figure}	

Here we show, numerically, that quantum calculations of YbTO favor 
the picture of a ferromagnet (FM) at low temperature. 
Using the finite temperature Lanczos method (FTLM)~\cite{Prelovsek,Bronca_review}, 
on an effective spin 1/2 anisotropic model on the pyrochlore lattice, 
we find good agreement with the experimentally observed Schottky anomaly centered at $2.4$ K, 
and the approximate location of the transition at low temperature~\cite{Coldea2017,Scheie2017,Arpino2017,Viviane2017}. 
Our approach complements previous reports~\cite{Applegate, Hayre} on YbTO with the numerical linked cluster (NLC) method. 
In addition, our calculations in a [111] magnetic field indicate that YbTO has 
substantial magnetization at small field strengths.

\begin{figure*}[htpb]
\centering
\includegraphics[width=0.325\linewidth]{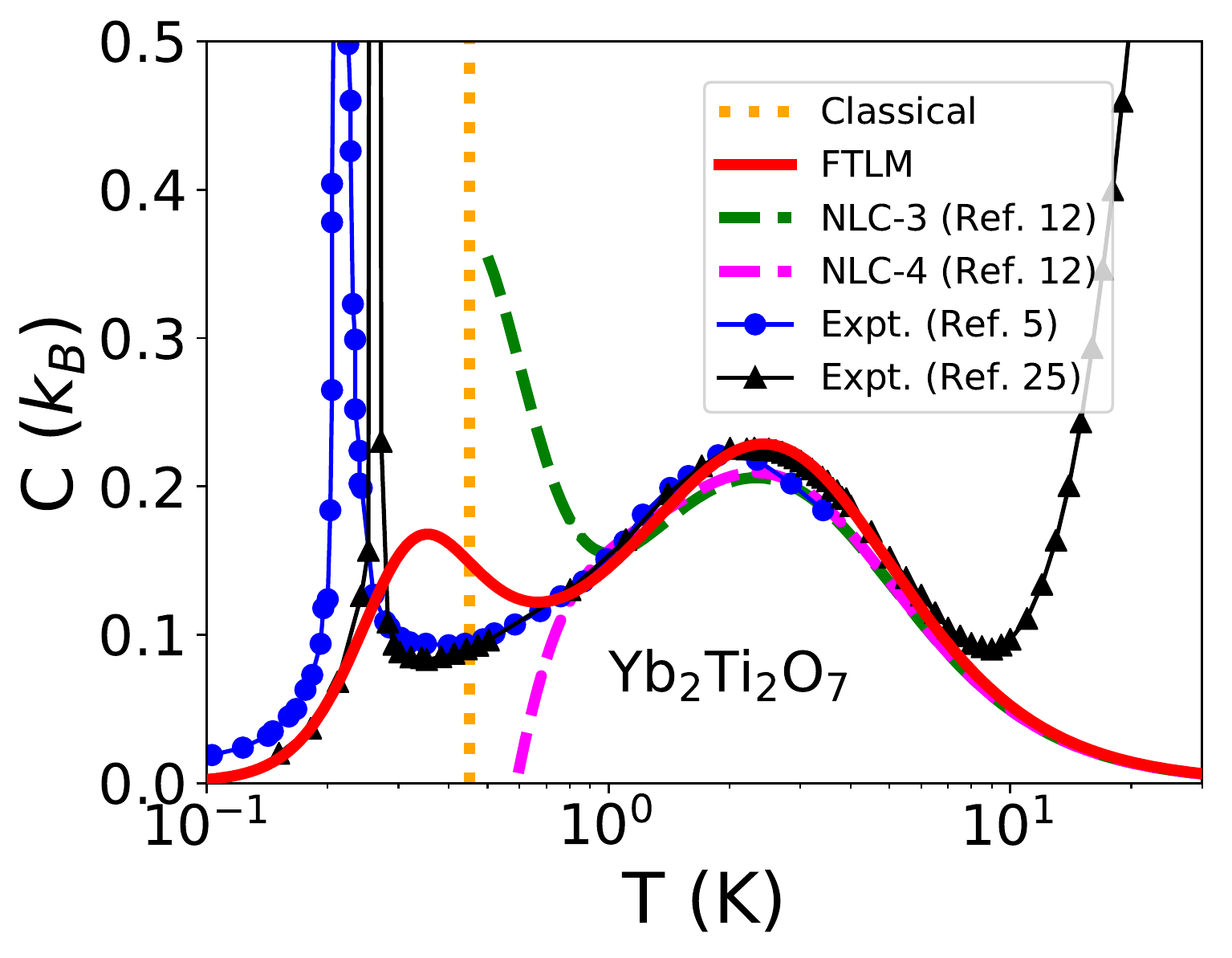}
\includegraphics[width=0.325\linewidth]{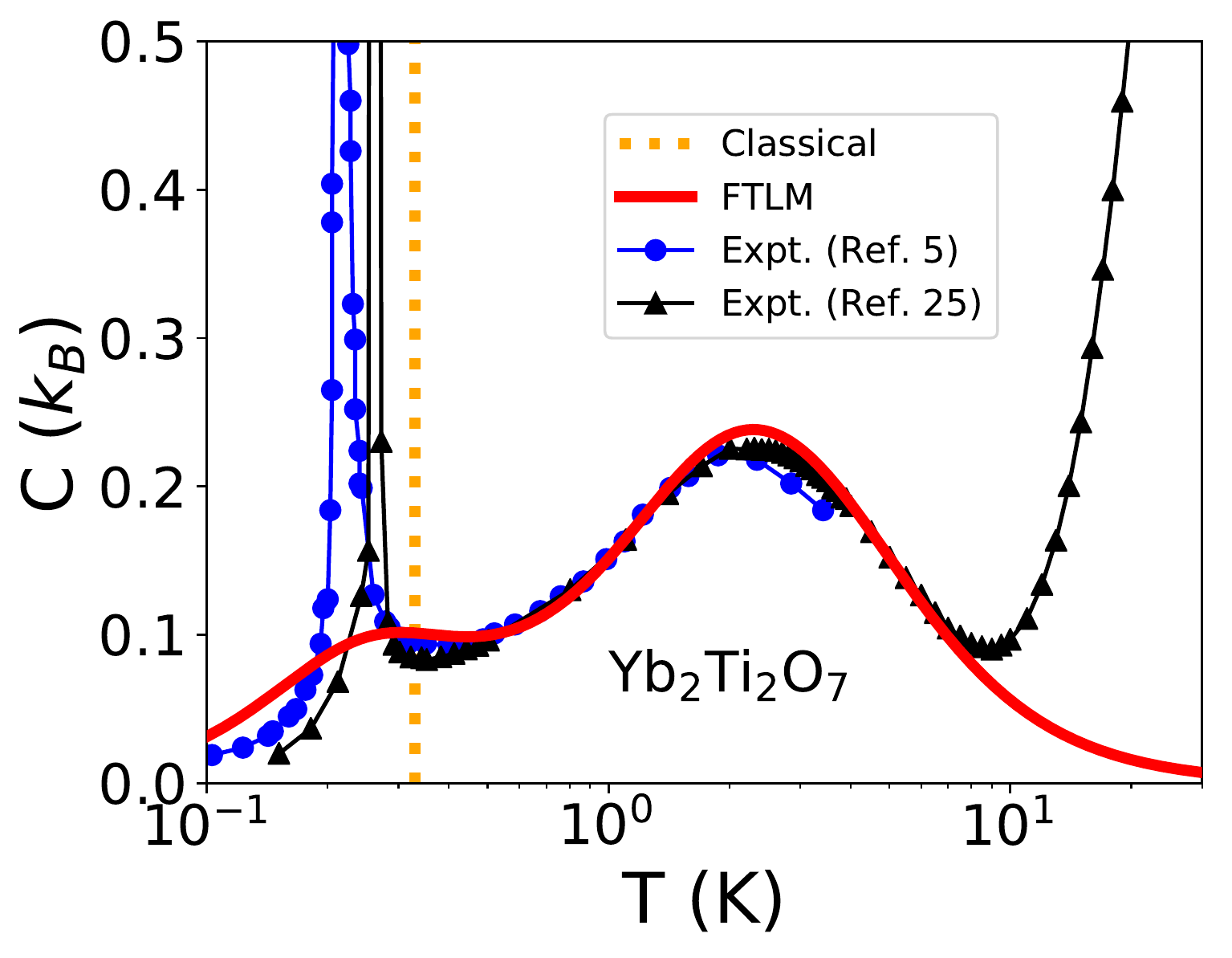}
\includegraphics[width=0.335\linewidth]{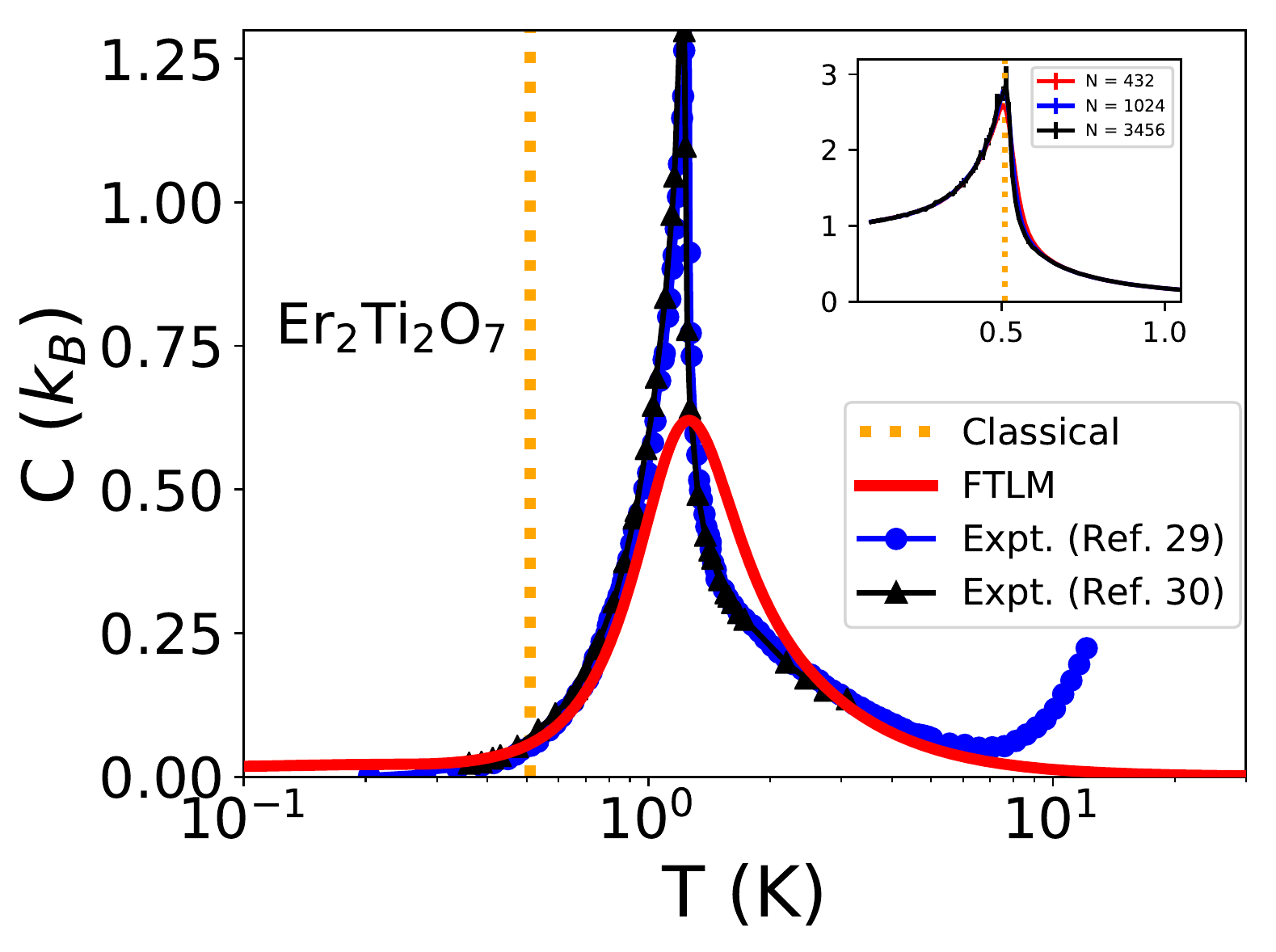}
\caption{(Color online): Heat capacity (per mole of the magnetic ion) of Yb$_2$Ti$_2$O$_7$ 
and Er$_2$Ti$_2$O$_7$ in units of the Boltzmann constant ($k_B$) as a function of temperature. 
The theoretical FTLM calculations on the $32$ site cluster 
used model parameters from Ref.~\cite{Ross_PRX} and Ref.~\cite{Coldea2017} for Yb$_2$Ti$_2$O$_7$ 
(left and central panels respectively) and Ref.~\cite{Savary_ErTO_order_disorder} for Er$_2$Ti$_2$O$_7$ (right panel). 
For comparison, results from experimental data sets~\cite{Bloete,Arpino2017,Dalmas,Niven} and previous NLC 
calculations~\cite{Applegate} are shown. The classical estimates of $T_c$ are obtained with classical Monte Carlo, 
and the results of the profile for Er$_2$Ti$_2$O$_7$ are shown in the inset.}
\label{fig:rossCvsT} 
\end{figure*}	

We discuss three main results. First, we demonstrate that quantum effects are crucial for the finite 
temperature properties of YbTO even for temperatures greater than the ordering temperature $T_c$; 
frustrated interactions and quantum effects renormalize $T_c$ significantly. 
This complexity is manifest in a magnetic field, and responsible for its unusual reentrant lobed 
phase diagram~\cite{Scheie2017}, thus, an explanation of recent experiments~\cite{Scheie2017,Coldea2017,Viviane2017} 
constitute the second purpose of this study. Finally, we present a simple picture in parameter space connecting 
the ice manifold to the YbTO parameter set, suggesting that the FM obtained within perturbation theory is 
connected to its counterpart in the non-perturbative regime.  
For these purposes, we have carried out simulations on $32$ sites as in Fig.~\ref{fig:32site}(a), 
with a Hilbert space per momentum sector of approximately $536$ million, 
significantly larger than previous quantum treatments~\cite{Onoda_quantum,Jaubert_multiphase} on the same model. 

The relevant effective spin Hamiltonian including the nearest neighbor interactions and onsite 
Zeeman coupling to an external magnetic field ($h=(h_x,h_y,h_z)$) is~\cite{Curnoe,Onoda_2011,Ross_PRX},
\begin{equation}
H = \frac{1}{2} \sum_{ij} J^{\mu\nu}_{ij} S^{\mu}_{i} S^{\nu}_{j} - \mu_{B} h^{\mu} \sum_{i} g^{\mu \nu}_{i} S^{\nu}_{i}
\label{eq:Ham}
\end{equation}
where $i,j$ are nearest neighbors and $\mu,\nu$ refer to $x,y,z$, $S^{\mu}_i$ refer to the spin $1/2$ components at site $i$, and 
 $\bf{J_{ij}}$ and $\bf{g_i}$ are bond and site dependent interaction and coupling matrices respectively, the former characterized completely 
by four
and the latter by two independent parameters.
The interaction part is most illuminating when written in terms of spin directions along the local [111] axes~(denoted by $\Scal$),
\begin{eqnarray}
H_{int} &=& \sum_{\langle i,j \rangle}   (2-\lambda) J_{zz} \;\; \Scal^{z}_{i}\Scal^{z}_{j}- \lambda J_{\pm} \Big( \Scal^{+}_{i} \Scal^{-}_{j} + \Scal^{-}_{i} \Scal^{+}_{j} \Big) \nonumber \\ 
  & &		  		+ \lambda J_{\pm \pm}\;\;\Big( \gamma_{ij} \Scal^{+}_{i} \Scal^{+}_{j} + \gamma^{*}_{ij} \Scal^{-}_{i} \Scal^{-}_{j} \Big) \nonumber \\ 
  & &	+ \lambda J_{z,\pm} \;\; \Big[ \Scal^{z}_{i} \Big( \Scal^{+}_{j} \zeta_{ij} + \Scal^{-}_{j} \zeta^{*}_{ij} \Big) + i \leftrightarrow j \Big] \label{eq:int_local}
\end{eqnarray}
where $J_{zz},J_{\pm},J_{\pm\pm},J_{z,\pm}$ are couplings and the parameter $\lambda$ has been introduced 
to tune from the classical ice manifold ($\lambda=0$) to material-relevant parameters ($\lambda=1$). 
$\gamma_{ij}$ and $\zeta_{ij}$ are bond dependent phases, the corresponding $4\times4$ matrices 
have been written out in the supplementary information. 
We work with $\lambda=1$, unless otherwise noted. 

\prlsec{Classical versus quantum effects on the specific heat capacity in zero field} 
We now discuss the results of the temperature dependence of the specific heat of YbTO 
in zero field using FTLM. This Krylov space method constructs an effective Hamiltonian in the space of suitably 
chosen vectors (powers of $H$ on a random vector) and calculates observables (that commute with $H$) from it. 
The efficiency of FTLM crucially depends on the adequacy of a small number of powers of the Hamiltonian ($M$) 
and a small number of starting random vectors ($R$)~\cite{Prelovsek,Bronca_review,Hanebaum2014}. 
Some details of the method and its convergence properties are discussed in the supplement.

\begin{figure*}[htpb]
\centering
\includegraphics[width=0.47\linewidth]{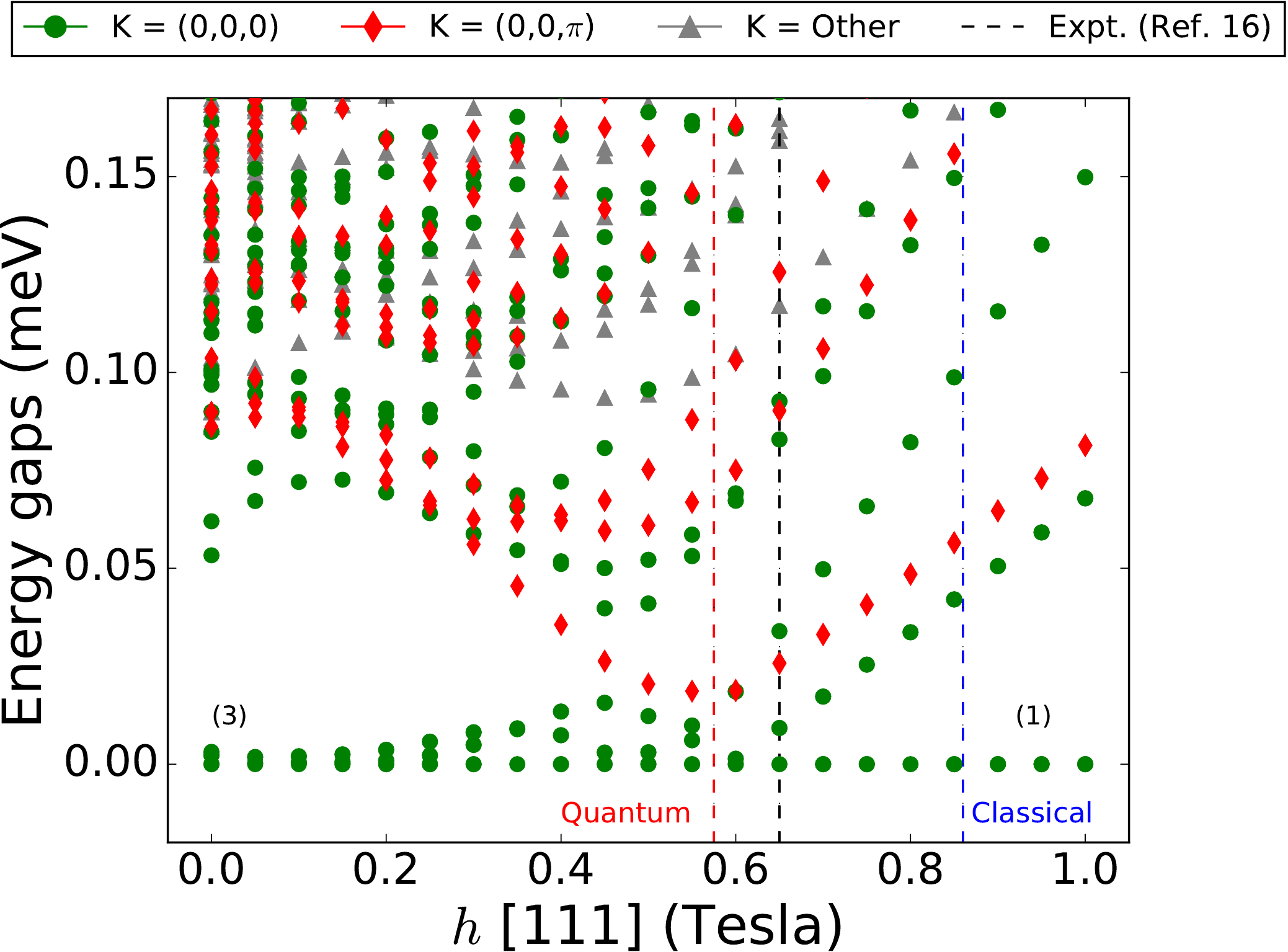}
\includegraphics[width=0.47\linewidth]{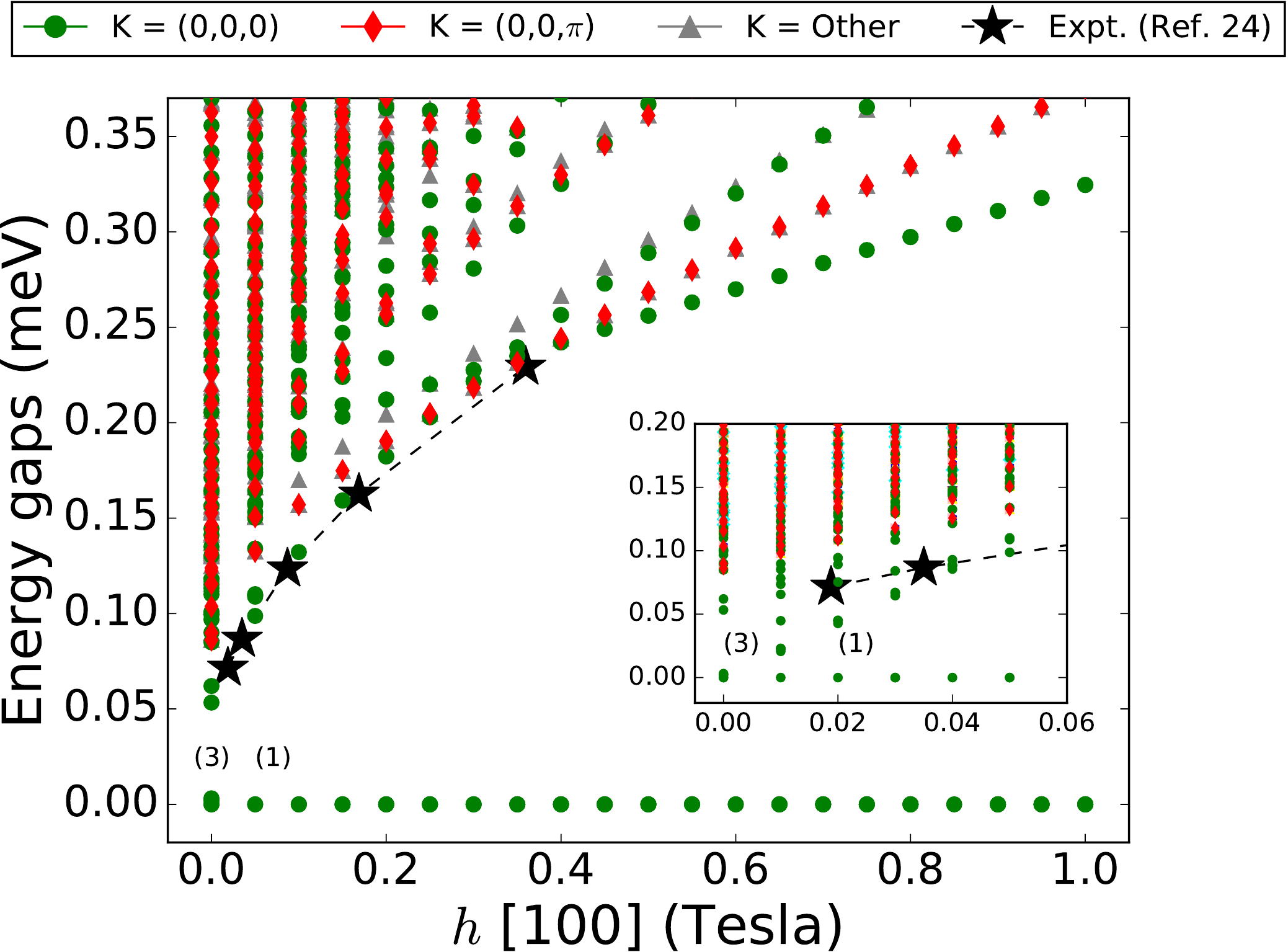}
\caption{(Color online): Energy (relative to the ground state) versus magnetic field ($h$) in [111] (left) and [100] (right) 
directions plotted by momentum sector ($k_1$,$k_2$,$k_3$). $M=400$ iterations were used for ${\bf k}=(0,0,0)$ 
and ${\bf k}=(0,0,\pi)$ and $M=200$ for the other sectors (denoted by $\bf k =$ other). 
$k_1$ and $k_2$ directions correspond to translations perpendicular to [111] 
and the $k_3$ direction corresponds to translations along [111]. The dashed lines 
indicate the classical, quantum and experimental estimates of the critical field at zero temperature. 
There is no phase transition for the [100] direction, although there is reorganization of the energy levels 
at small fields (inset). The experimental field dependent gap is also shown for the [100] case. For both cases, 
the quasidegeneracies of the ground state are shown in parentheses.}
\label{fig:zeroTfinitefield} 
\end{figure*}	

The leftmost panel of Fig.~\ref{fig:rossCvsT} shows the converged specific heat profile (for the parameters of Ref.~\cite{Ross_PRX}) 
across a wide range of temperatures for $M=100$ and $R=160$, along with two experimental data sets~\cite{Arpino2017,Bloete}. 
The agreement with the NLC approach at high temperatures is excellent, 
which serves as a check of our calculations. Importantly, the quantum treatment yields a small "peak" ($T_c\approx 0.34$ K), 
which appears as a crossover, but is not accessible in the NLC approach~\cite{Applegate}. When compared 
to classical Monte Carlo (MC) simulations, which yield $T_c \approx 0.45$ K, we conclude that $T_c$ is renormalized due to quantum 
effects ($0.11$ K is not a small scale for the phenomenon relevant to experiments, as will become clear shortly). 
Based on finite size analyses (of the MC calculations), the extent of the change in $T_c$ on approaching the thermodynamic limit (TDL) 
is insufficient to reconcile the classical and quantum estimates. 

Prominently, the Schottky anomaly at $2.4$ K highlights the importance of quantum effects, 
since it is \emph{completely} absent from classical simulations in zero field. The 
agreement of this feature with experiments is remarkable; the deviations are small and not 
visualized on the scale of the plot. Even below $T_c$, the numerically computed values essentially lie on top of the experimental data. 
Above 10 K the experimental data includes contributions from non-magnetic degrees of freedom,
not part of the model Hamiltonian.

Recently, different parameters (with reduced $J_{zz}$) have been reported by Ref.~\cite{Coldea2017}; 
the central panel of Fig.~\ref{fig:rossCvsT} shows our results for this set. The 
Schottky anomaly is explained well here too, but importantly a lower $T_c$ is observed, 
both classically and quantum mechanically; the latter appears as a broad hump at $T_c \approx 0.27$ K. 
This suggests longer correlation lengths leading to more pronounced finite size effects and 
YbTO's possible closeness to a phase boundary~\cite{Jaubert_multiphase, Robert}. 
However, several aspects of experiments can be understood from the set of Ref.~\cite{Ross_PRX}
~(with smaller finite size effects); we use those for the remainder of the paper. 

While YbTO is the main focus of this work, we demonstrate the effectiveness of the approach for another 
pyrochlore, Er$_2$Ti$_2$O$_7$ (ErTO). ErTO is a candidate for the "order by disorder 
effect"~\cite{Henley_order_by_disorder,Villain} and has been extensively 
studied~\cite{Savary_ErTO_order_disorder,Gingras_ErTO,Moessner_ErTO,Gaulin_review}. 
We use the Hamiltonian parameters from Ref.~\cite{Savary_ErTO_order_disorder}, and compare 
to two experimental data sets~\cite{Dalmas,Niven}, our results are 
in the rightmost panel of Fig.~\ref{fig:rossCvsT}. Unlike YbTO, ErTO displays no prominent Schottky anomaly~\cite{Gaulin_review}, 
but instead a single phase transition at $T_c=1.23\pm0.01$ K. 
The quantum calculations capture this effect, and we find $T_c \approx 1.26 $ K.
In contrast, the MC calculations (see inset for profiles for of systems ranging from $432$ to $3456$ sites) 
show a much lower $T_c \approx 0.51(1) $ K. 
\begin{figure}[htpb]
\centering
\includegraphics[width=0.9\linewidth]{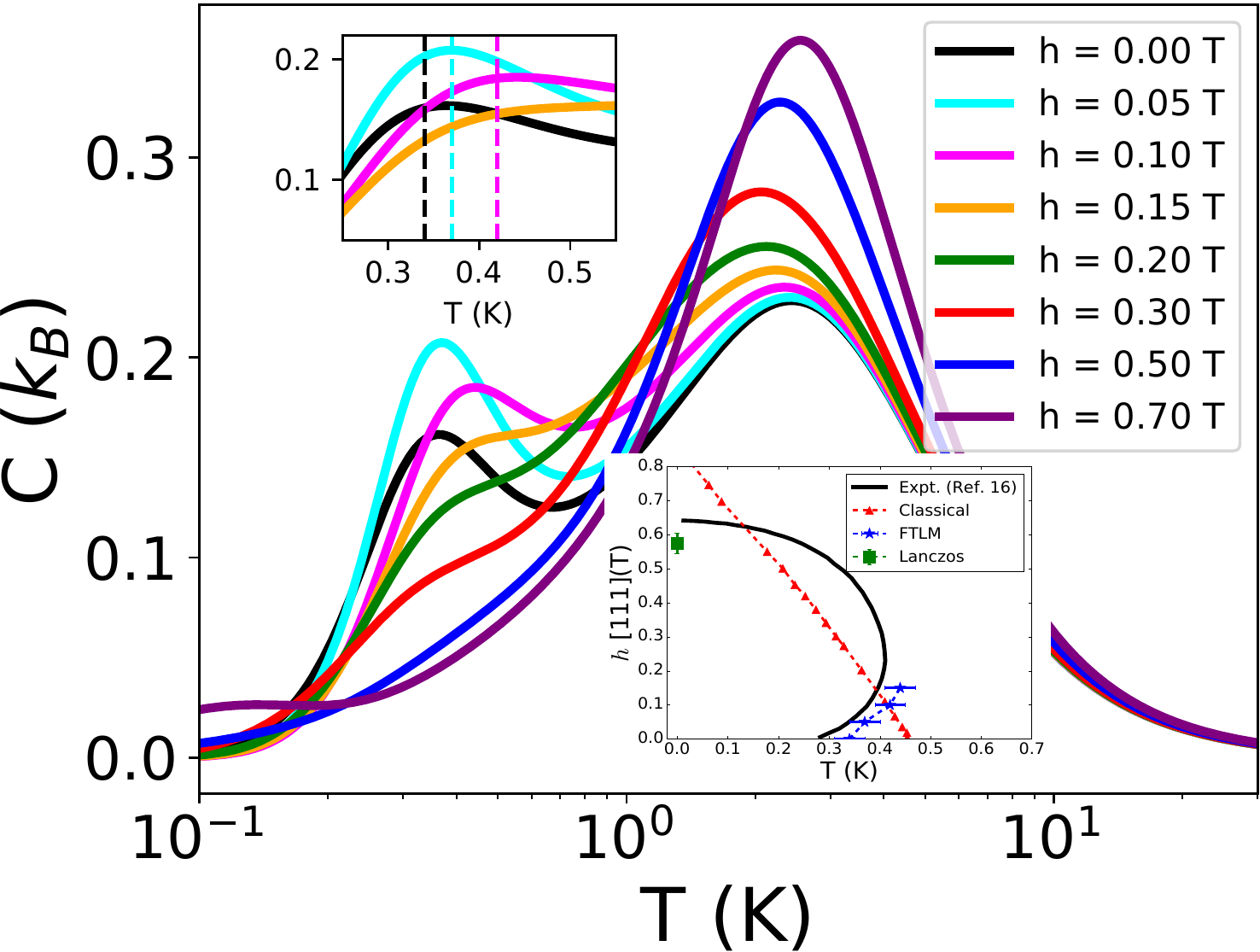}
\includegraphics[width=0.325\linewidth]{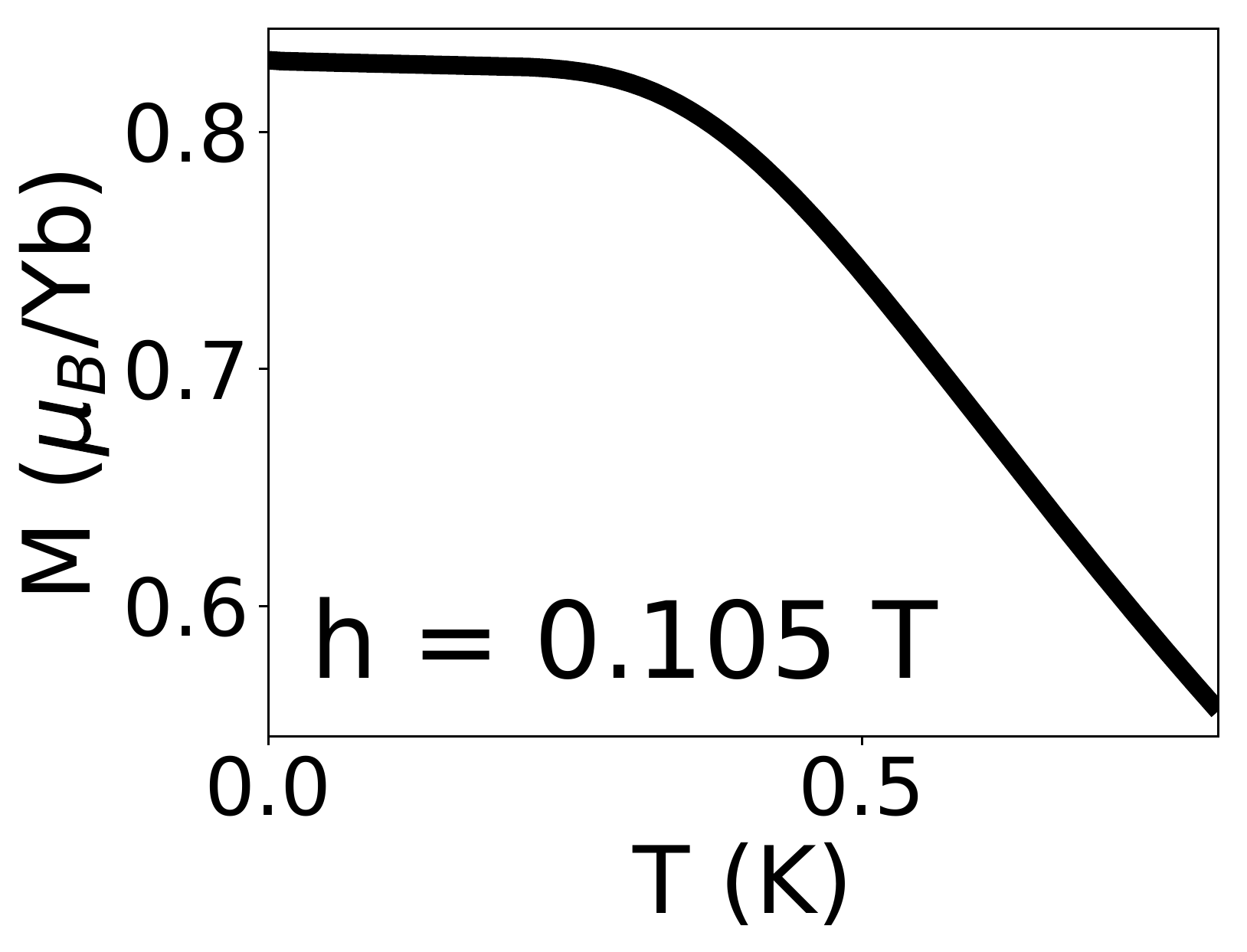}
\includegraphics[width=0.325\linewidth]{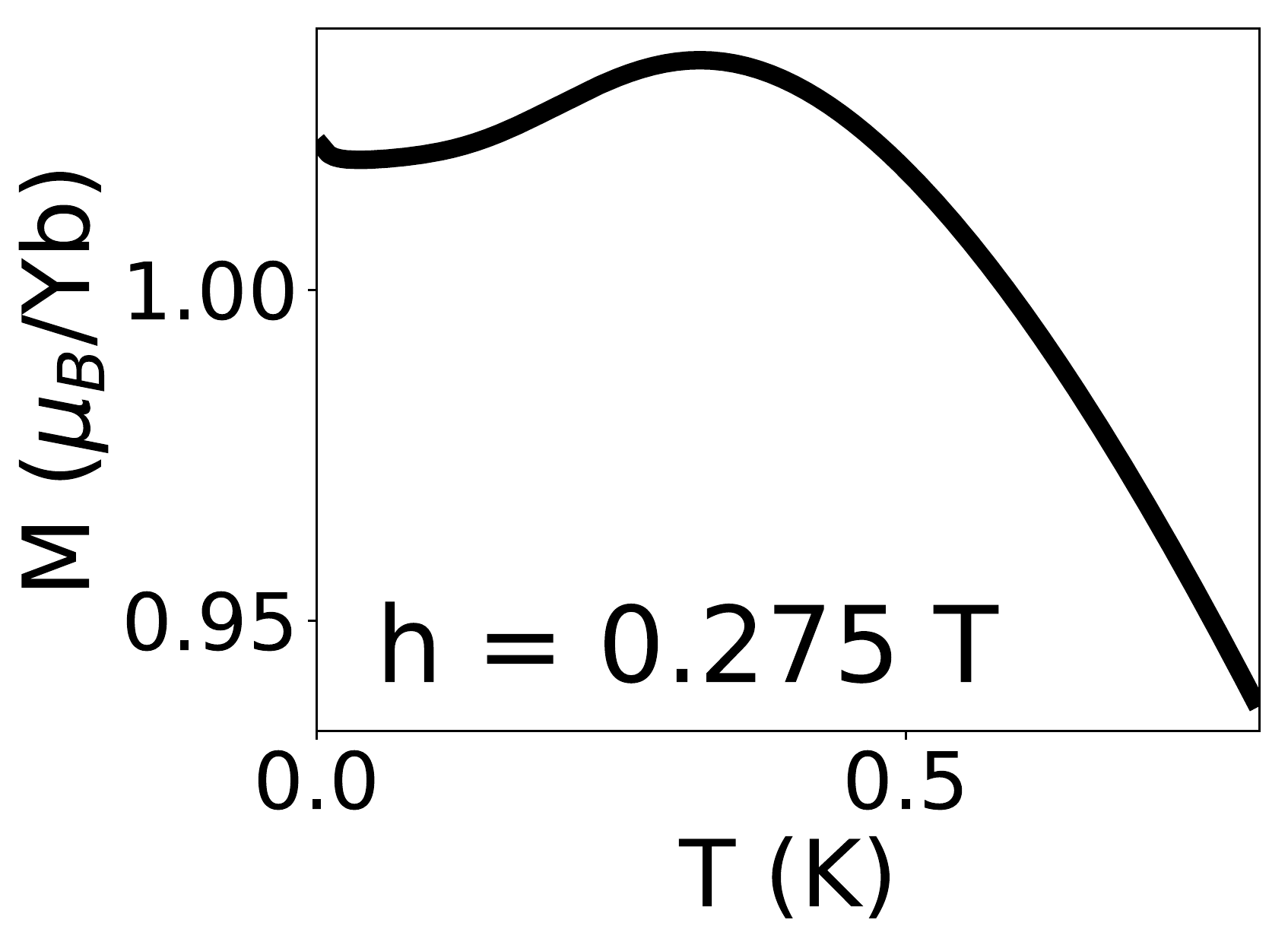}
\includegraphics[width=0.325\linewidth]{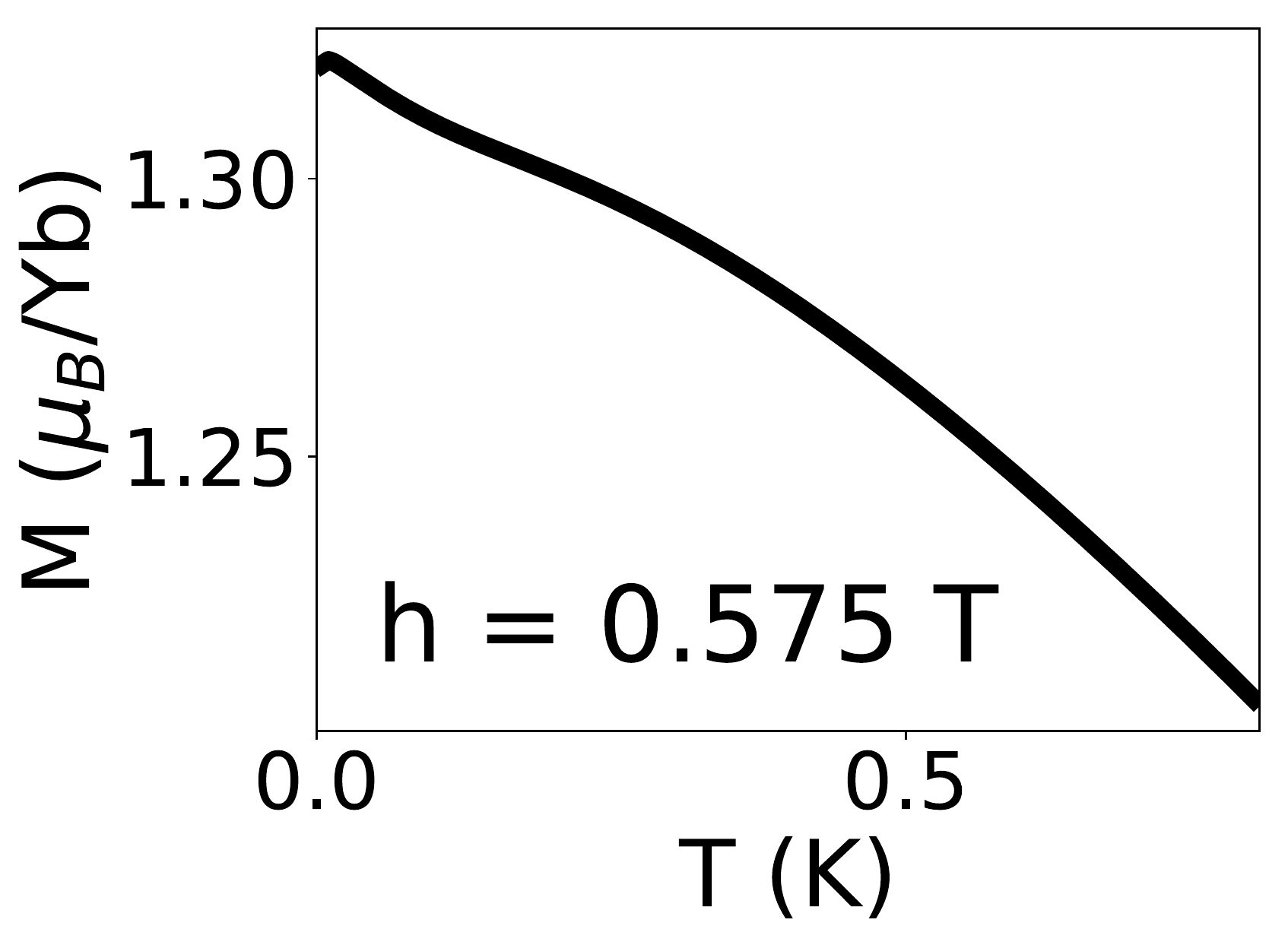}
\caption{(Color online): The top panel shows the heat capacity (per mole of Yb) in units of the Boltzmann constant ($k_B$) 
as a function of temperature for different [111] magnetic field strengths ($h$). The 
upper inset zooms in on the low temperature peak; the dashed lines indicate the location of $T_c$. (For $h \gtrsim 0.15 $T, the peaks are not visible). 
The lower inset shows the experimental lobed phase diagram and the quantum and classical 
phase boundaries. The lower panels show the magnetization ($M$) measured along the [111] 
direction as a function of temperature for various representative $h$.}
\label{fig:rossCvsTh} 
\end{figure}	

\prlsec{Low energy spectra and quantum phase transitions in a [111] and [100] magnetic field} 
The difference between classical and quantum treatments of YbTO is particularly striking at zero temperature, 
in a [111] magnetic field. Since the Hamiltonian has translational symmetry, all 
eigenstates have definite momenta ${\bf k} =(k_1,k_2,k_3)$; for the $32$ site cluster each $k_i=0$ or $k_i=\pi$. 
The lowest lying energies are mapped out in all $8$ momentum sectors in Fig.~\ref{fig:zeroTfinitefield}. Since only momenta ${\bf k}=(0,0,0)$ and ${\bf k}=(0,0,\pi)$ 
are involved in the zero temperature phase transition, their corresponding labels have been highlighted, the rest 
are denoted as "${\bf k}=$other".

In zero field there are three quasidegenerate states, all in ${\bf k}=(0,0,0)$ 
followed by two more states separated by approximately $0.05$ meV. This is at odds with 
six quasidegenerate states expected of a cubic FM; this can be attributed to the \emph{lack} of cubic symmetry 
of the $32$ site cluster and the large splittings between states. An analogous effect involving large 
tunneling between time reversal symmetric odd and even states is also seen on the kagome~\cite{Kumar_Changlani}. 
For finite fields, only three quasidegenerate states in ${\bf k}=(0,0,0)$ remain part of the 
low energy manifold while the other states separate out from these. 
(In the TDL, an infinitesimally small [111] field would gap out three of the six states for a cubic FM). 
Simultaneously, the lowest ${\bf k}=(0,0,\pi)$ state is lowered in energy. 
Then between $0.55$ to $0.60$ T, the lowest ${\bf k}=(0,0,\pi)$ state makes its closest approach to the quasidegenerate manifold; 
at this critical field ($h_c$), the ${\bf k}=(0,0,0)$ states also split into three branches. 
The observed $h_c$ agrees well with the experimental value ($0.65$ T)~\cite{Scheie2017}, and is significantly lower than the classical estimate of $0.86(1)$ T. 

In contrast, in a [100] field (right panel of Fig.~\ref{fig:zeroTfinitefield}), 
there is no sign of a phase transition, consistent with the findings of Ref.~\cite{Coldea2017}. 
At small fields ($<0.03 $T), the three quasidegenerate states split (see inset) yielding 
a single non-degenerate ground state which is qualitatively the same as the high field limit. 
The trends in the excited eigenenergies also explain the field dependent gap of Ref.~\cite{Coldea2017}. 
These observations collectively suggest YbTO is a FM. 

\prlsec{Specific heat and magnetization in a [111] magnetic field} 
It is now natural to ask how quantum fluctuations manifest themselves at finite field and 
temperature. Recently, Ref.~\cite{Scheie2017} found that on increasing the [111] field from $0$ to $0.2$ T, 
$T_c$ increased significantly from $0.27$ K to $0.42 $K, before it gradually decreased towards zero for 
higher fields, yielding a reentrant lobed phase diagram. While this initial 
increase is expected of a first order phase transition, the \emph{magnitude} of this effect \emph{could not} 
be captured classically. 

To address this issue, we performed FTLM calculations in a [111] field $h$; our results are shown in Fig.~\ref{fig:rossCvsTh}. 
(As a compromise between accuracy and computer time, we chose $R=40$ and $M=50$). On increasing $h$, 
the Schottky anomaly increases in height and the associated entropy increases. Quantum mechanically, 
the combined entropy of the peak ($S_{peak}$) and anomaly must be constant ($\ln (2)$ in units of $k_B$, ignoring non-magnetic contributions), 
implying $S_{peak}$ decreases with increasing $h$. Next, the upper inset of Fig.~\ref{fig:rossCvsTh} 
shows the low temperature peaks, marked by dashed lines, the broad features they are associated with move right by
$\approx 0.1 $ K. Both observations agree with experimental findings~\cite{Scheie2017}; 
the lower inset compares their phase diagram and our simulations.

These inferences are confirmed by the magnetization along the [111] direction ($M$)
\begin{equation}
	M (T,h) = k_B T \frac{\partial \ln Z(T,h)} {\partial h} 
\end{equation}
where $Z(T,h)$ is the partition function directly calculated in FTLM. 
$M(T,h)$ was evaluated using finite differences; $\delta h$ was chosen to be $0.01$ T for small fields, and $0.05$ T for larger fields.
Representative results are shown in Fig.~\ref{fig:rossCvsTh}. 

$M$($T\rightarrow0,h$) is non zero for small $h$ and increases with 
$h$ suggesting that YbTO is a FM at low temperatures; noting that the distinction between a FM and very good paramagnet (PM) 
is difficult for a $32$ site system (also see supplement) In addition, the quantum simulations show signs of 
a phase transition from FM to PM, for example 
a (smooth) kink is seen in $M(T,h)$ up to $\approx 0.15$ T. Then for intermediate $h$
($h\approx 0.15$ T to $h \approx 0.5$ T), $M$ in the putative FM phase is lower than that in the PM, 
which is consistent with $T_c$ decreasing with increasing $h$. By $0.55$ T, there is no phase transition, 
consistent with the zero temperature findings. 
\begin{figure}[htpb]
\centering
\includegraphics[width=0.9\linewidth]{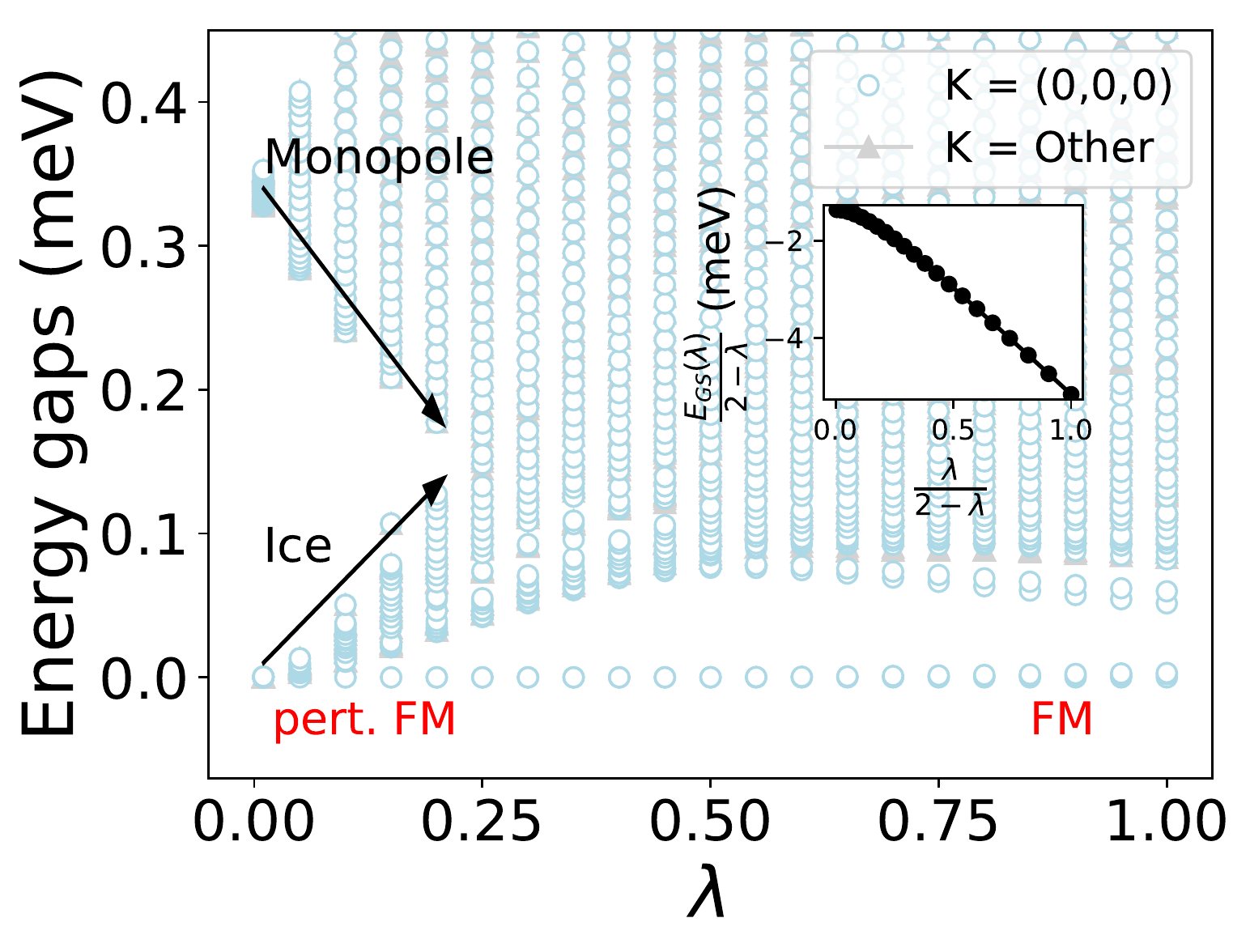}
\caption{(Color online): Energy (relative to the ground state) versus $\lambda$, the 
parameter defined in the text, that allows us to tune from the classical ice manifold ($\lambda=0$) 
to the experimentally relevant parameters ($\lambda=1$). 
The inset shows the total ground state energy which smoothly evolves from the perturbative to non-perturbative 
regime.}
\label{fig:zeroTlambda} 
\end{figure}	

\prlsec{Relation of YbTO to the ice manifold} 
We now provide a qualitative interpretation of the existence of an FM by varying 
$\lambda$ (in Eq.~\eqref{eq:int_local}) from $0$ (classical ice) to $1$ (YbTO-relevant parameters) and monitoring the spectrum, our results are 
shown in Fig.~\ref{fig:zeroTlambda}. When $\lambda$ is small, perturbative arguments apply due to 
presence of the gap of order $(2-\lambda)J_{zz}$. The leading order contribution is to second order; 
the $J_{z\pm}$ term selects the six FM states~\cite{Ross_PRX,Savary_gaugeMFT}. 
When $\lambda$ is increased, the energy of the monopole (defect) manifold decreases relative to the 
highest energy of the ice-like manifold, which itself has split,
and around $\lambda \approx 0.25$ the two features meet. However, the ground state is qualitatively unaffected 
by this high energy feature as the state selection from the ice manifold has already occurred at a much lower energy scale 
(the inset shows the ground state energy evolving from the perturbative regime to the non-perturbative one). 
Thus, we suggest that the FM seen in perturbation theory is connected to the FM established earlier in the paper, 
for the parameters of Ref.~\cite{Ross_PRX}. This adiabatic connection suggests YbTO is a near-collinear FM~\cite{Chang_Onoda,Yasui}. 

\prlsec{Conclusion} In summary, quantum mechanical effects are crucial to understanding 
the zero and finite temperature properties of YbTO. 
Despite the limitations of finite size ($32$ sites), we observed good agreement between available experimental data and simulations 
for a wide range of temperatures. In finite [111] fields, we saw a significant \emph{increase} 
in $T_c$ of $\approx 0.1$ K till $\approx 0.15$ T. The presence of substantial 
magnetization at small fields suggests YbTO is a FM. This picture is strengthened by connecting our results 
to those known within second order perturbation theory of the ice manifold. 

Finally, we showed the utility of FTLM for large Hilbert spaces, for meaningfully 
studying certain properties of other pyrochlores at finite temperature, 
as long as the correlation lengths are sufficiently short. One hopes that recent numerical advances 
at zero and finite temperature~\cite{White_finite_temp, Claes_Clark, Changlani_CPS, 
Holmes_heatbath, Semistoch_Petruzielo, Holmes_HCI, Tubman_selci, Blunt_Alavi} will 
push the limits of interesting sizes (of high-dimensional systems) one can go to with some degree of confidence. 

\prlsec{Acknowledgements} I thank O. Tchernyshyov and S. Zhang for teaching me 
the basics of this field of research and for extensive discussions. I gratefully acknowledge 
C. Broholm, A. Scheie, J. Kindervater, 
R.R.P. Singh, R. Coldea, M. Gingras, J. Rau, Y. Wan, 
K. Plumb, S. S\"aubert, A. Eyal, S. Koohpayeh, L. Jaubert and B. Clark for useful discussions and D. Kochkov 
for sharing large-scale exact diagonalization tricks. 
I thank R.R.P. Singh for sharing his NLC data and for critically reading the first draft of this manuscript, 
B. Gaulin, A. Hallas and J. Gaudet for providing experimental specific heat data sets from their review article, and 
A. Scheie and K. Arpino for providing data from their published work. 
This work was supported through the Institute for
Quantum Matter at Johns Hopkins University, by the
U.S. Department of Energy, Division of Basic Energy
Sciences, Grant DE-FG02-08ER46544. I gratefully acknowledge the Johns 
Hopkins Homewood High Performance Cluster (HHPC) and the Maryland Advanced Research Computing Center (MARCC), 
funded by the State of Maryland, for computing resources and Yi Li for sharing part of her computer 
allocation time.

\bibliographystyle{apsrev4-1}
\bibliography{refs}

\pagebreak
\textbf{\large Supplemental Material for "Quantum versus classical effects at zero and finite temperature in the quantum pyrochlore Yb$_2$Ti$_2$O$_7$"} 

\setcounter{equation}{0}
\setcounter{figure}{0}
\setcounter{table}{0}
\setcounter{page}{1}
\makeatletter
\renewcommand{\theequation}{S\arabic{equation}}
\renewcommand{\thefigure}{S\arabic{figure}}
\renewcommand{\bibnumfmt}[1]{[#1]}
\renewcommand{\citenumfont}[1]{#1}

\maketitle
\section{Low energy effective Hamiltonian}
In this section, we discuss details of the relevant spin $1/2$ low-energy effective Hamiltonian on the pyrochlore lattice, 
with nearest neighbor interactions and Zeeman coupling to an external field ($h=(h_x,h_y,h_z)$)~\cite{Curnoe,Onoda_2011,Ross_PRX},
\begin{equation}
H = \frac{1}{2} \sum_{ij} J^{\mu\nu}_{ij} S^{\mu}_{i} S^{\nu}_{j} - \mu_{B} h^{\mu} \sum_{i} g^{\mu \nu}_{i} S^{\nu}_{i}
\label{eq:Ham_supp}
\end{equation}
where $i,j$ are nearest neighbors and $\mu,\nu$ refer to $x,y,z$, $S^{\mu}_i$ refer to the spin components at site $i$, and 
 $\bf{J}_{ij}$ and $\bf{g}_i$ are bond and site dependent interactions and coupling matrices respectively 
(whose components have been written out in Eq.~\ref{eq:Ham_supp}). The pyrochlore lattice 
has four sublattices which we label as $0,1,2,3$ and we take the relative locations of the sites on a single tetrahedron to be, 
(in units of lattice constant $a$) ${\bf{r}_0}=(1/8,1/8,1/8)$, ${\bf{r}_1}=(1/8,-1/8,-1/8)$, 
${\bf{r}_2}=(-1/8,1/8,-1/8)$ and ${\bf{r}_3}=(-1/8,-1/8,1/8)$. Symmetry considerations dictate that $\bf{J}_{ij}$ and $\bf{g}_{i}$ are 
completely described by four and two scalars respectively. 
$\bf{J}_{ij}$ depends only on the sublattices that $i,j$ belong to
(similarly $\bf{g}_i$ depends only on the sublattice of site $i$), and thus we use the notation 
in terms of $i,j=0,1,2,3$. Also, since $\bf{J}_{ij}=\bf{J}_{ji}^T$, only the $i<j$ matrices are 
written out. The $\bf{J}_{ij}$ matrices are,
\begin{equation}
\bf{J}_{01} \equiv
\left(\begin{array}{ccc}
 J_2 & J_4 & J_4 \\
-J_4 & J_1 & J_3 \\
-J_4 & J_3 & J_1 \end{array} \right) 
\bf{J}_{02} \equiv
\left(\begin{array}{ccc}
 J_1 & -J_4 & J_3 \\
 J_4 & J_2 & J_4 \\
 J_3 & -J_4 & J_1 \end{array} \right) \nonumber
\end{equation}
\begin{equation}
\bf{J}_{03} \equiv
\left(\begin{array}{ccc}
 J_1 & J_3 & -J_4 \\
 J_3 & J_1 & -J_4 \\
 J_4 & J_4 & J_2 \end{array} \right) 
\bf{J}_{12} \equiv
\left(\begin{array}{ccc}
 J_1 & -J_3 & J_4 \\
-J_3 & J_1 & -J_4 \\
-J_4 & J_4 & J_2 \end{array} \right) \nonumber 
\end{equation}
\begin{equation}
\bf{J}_{13} \equiv
\left(\begin{array}{ccc}
 J_1 & J_4 & -J_3 \\
-J_4 & J_2 & J_4 \\
-J_3 & -J_4 & J_1 \end{array} \right) 
\bf{J}_{23} \equiv
\left(\begin{array}{ccc}
 J_2 & -J_4 & J_4 \\
 J_4 & J_1 & -J_3 \\
-J_4 & -J_3 & J_1 \end{array} \right) 
\end{equation}
Defining $g_{+}=\frac{1}{3}(2g_{xy}+g_{z})$ and $g_{-}=\frac{1}{3}(g_{xy}-g_z)$, the 
$\bf{g}_i$ matrices read as,
\begin{equation}
\bf{g}_{0} \equiv
\left(\begin{array}{ccc}
 g_+ & -g_- & -g_- \\
 -g_- & g_+ & -g_- \\
 -g_- & -g_- & g_+ \end{array} \right) 
\bf{g}_{1} \equiv
\left(\begin{array}{ccc}
 g_+ & g_- & g_- \\
 g_- & g_+ & -g_- \\
 g_- & -g_- & g_+ \end{array} \right) \nonumber
\end{equation}
\begin{equation}
\bf{g}_{2} \equiv
\left(\begin{array}{ccc}
 g_+ & g_- & -g_- \\
 g_- & g_+ & g_- \\
-g_- & g_- & g_+ \end{array} \right) 
\bf{g}_{3} \equiv
\left(\begin{array}{ccc}
 g_+ & -g_- & g_- \\
-g_- & g_+ & g_- \\
 g_- & g_- & g_+ \end{array} \right) 
\end{equation}

The interaction part when written in terms of spin directions along the local [111] axes~(denoted by $\Scal$), is,
\begin{eqnarray}
H_{int} &=& \sum_{\langle i,j \rangle}   (2-\lambda) J_{zz} \;\; \Scal^{z}_{i}\Scal^{z}_{j}- \lambda J_{\pm} \Big( \Scal^{+}_{i} \Scal^{-}_{j} + \Scal^{-}_{i} \Scal^{+}_{j} \Big) \nonumber \\ 
  & &		  		+ \lambda J_{\pm \pm}\;\;\Big( \gamma_{ij} \Scal^{+}_{i} \Scal^{+}_{j} + \gamma^{*}_{ij} \Scal^{-}_{i} \Scal^{-}_{j} \Big) \nonumber \\ 					 
  & &		+ \lambda J_{z,\pm} \;\; \Big[ \Scal^{z}_{i} \Big( \Scal^{+}_{j} \zeta_{ij} + \Scal^{-}_{j} \zeta^{*}_{ij} \Big) + i \leftrightarrow j \Big] \label{eq:Ham}
\end{eqnarray}
where $J_{zz},J_{\pm},J_{\pm\pm},J_{z,\pm}$ are couplings and the parameter $\lambda$ has been introduced by us 
to tune from the classical ice manifold ($\lambda=0$) to real material relevant parameters ($\lambda=1$). 
$\zeta_{ij}$ and $\gamma_{ij}$ are bond dependent phases, 
\begin{equation}
\zeta \equiv
\left(\begin{array}{cccc}
  0 & -1 & e^{+i\pi/3} & e^{-i\pi/3} \\
 -1 & 0  & e^{-i\pi/3} & e^{+i\pi/3} \\
 e^{+i\pi/3} & e^{-i\pi/3} & 0 & -1  \\
 e^{-i\pi/3} & e^{+i\pi/3} & -1 & 0 \end{array} \right) 
\gamma = \zeta^{*}
\end{equation}

The relation between $J_1,J_2,J_3,J_4$ and  $J_{zz},J_{\pm},J_{\pm\pm},J_{z,\pm}$ is, 
\begin{eqnarray}
J_{zz}     &=& -\frac{1}{3}         (+2 J_{1} - J_{2} + 2 J_{3} +  4 J_{4})     \nonumber \\ 
J_{\pm}    &=& +\frac{1}{6}         (+2 J_{1} - J_{2} -   J_{3} -  2 J_{4})     \nonumber \\ 
J_{\pm\pm} &=& +\frac{1}{6}         (+  J_{1} + J_{2} - 2 J_{3} +  2 J_{4})     \nonumber \\ 
J_{z\pm}   &=& +\frac{1}{3\sqrt{2}} (+  J_{1} + J_{2} +   J_{3} -    J_{4})     
\end{eqnarray}

We provide a table of the parameters that were used for the calculations in both notations. For Yb$_2$Ti$_2$O$_7$~(YbTO), 
parameters from Ref.~\cite{Ross_PRX} and Ref.~\cite{Coldea2017}, and for Er$_2$Ti$_2$O$_7$~(ErTO), 
parameters from Ref.~\cite{Savary_ErTO_order_disorder} were used.
The parameters from one notation are directly converted to the other notation (without accounting for error bars) 
unless already provided in the reference. 

\begin{table*}[htpb]
\begin{center}
\begin{tabular}{|c|c|c|c|c||c|c|c|c||c|c|}
\hline
Parameter set     &  $J_1$(meV)  & $J_2$ (meV) & $J_3$ (meV) & $J_4$ (meV) &  $J_{zz}$ (meV) & $J_{\pm}$ (meV) & $J_{z\pm}$ (meV) & $J_{\pm \pm}$ (meV) & $\;\;g_{xy}\;\;$ & $\;\;g_{z}\;\;$  \tabularnewline
\hline
YbTO Ref.~\cite{Ross_PRX}  &  -0.09 & -0.22 & -0.29 & +0.01 &  0.17   & 0.05   & -0.14 & 0.05  &  4.32   & 1.8    \tabularnewline
YbTO Ref.~\cite{Coldea2017}&  -0.028 & -0.326 & -0.272 & +0.049 &  0.026  & 0.074  & -0.159 & 0.048 &  4.17   & 2.14   \tabularnewline
ErTO Ref.~\cite{Savary_ErTO_order_disorder} & +0.115 & -0.056 & -0.099 & -0.003 &  -0.025    & 0.065 & -0.0088   & 0.042 &  5.97 & 2.45 \tabularnewline
\hline
\end{tabular}
\caption{Parameter sets used in the paper in two notations for YbTO and ErTO. The reported parameters 
from one notation are directly converted to the other notation (without accounting for error bars) 
unless already provided in the reference.}
\label{tab:parameters}
\end{center}
\end{table*}

\begin{figure*}[tpb]
\centering
\includegraphics[width=0.325\linewidth]{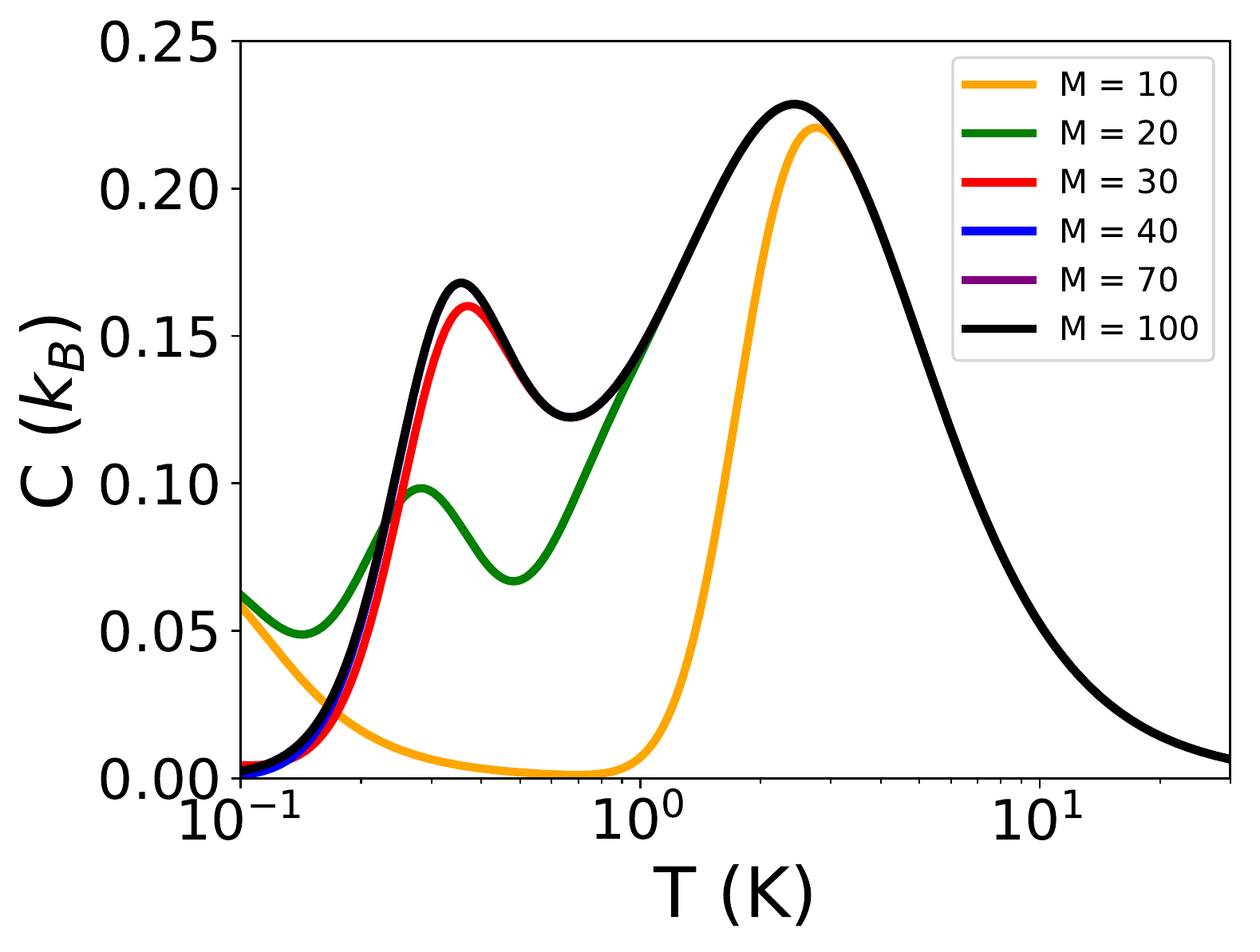}
\includegraphics[width=0.325\linewidth]{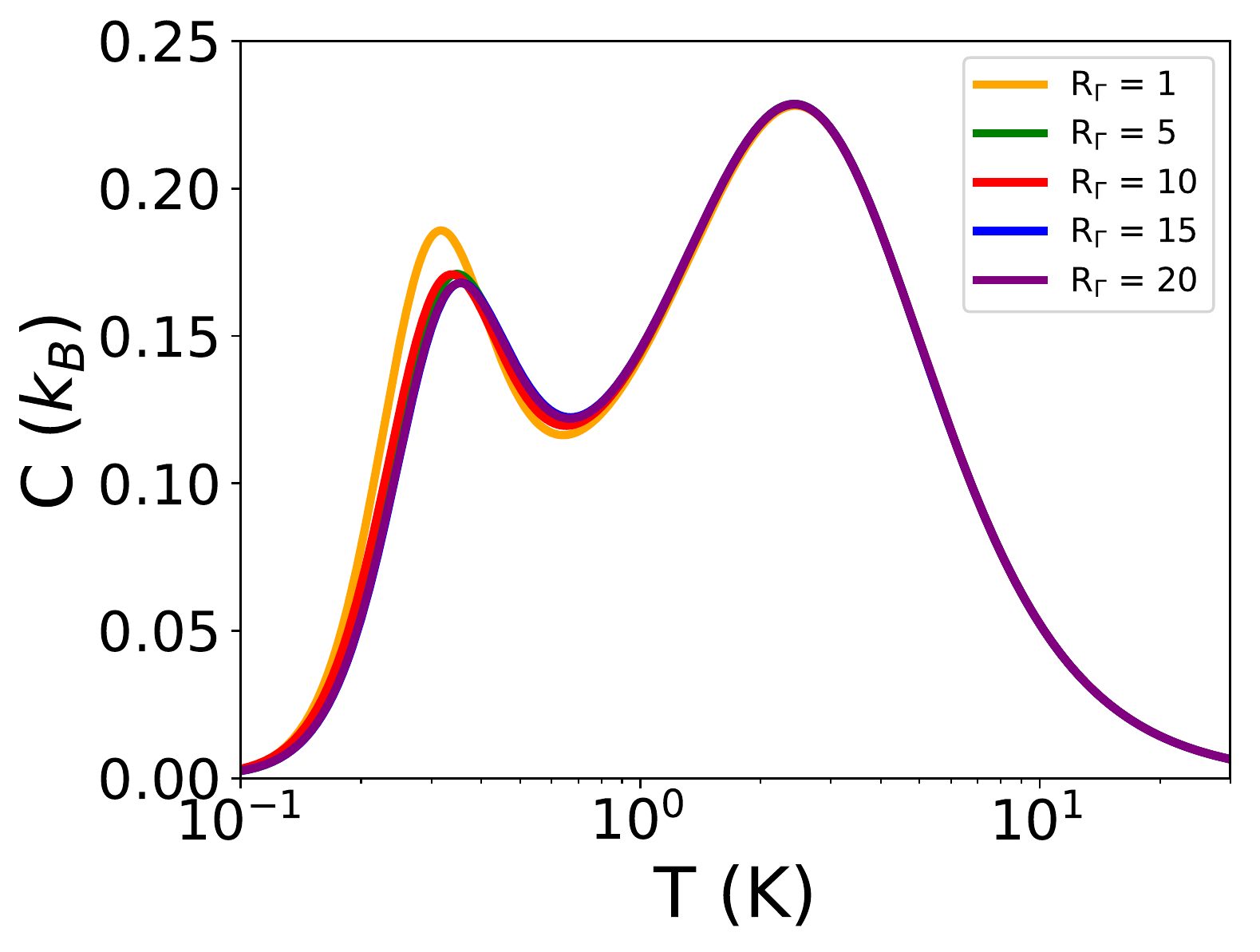}
\includegraphics[width=0.325\linewidth]{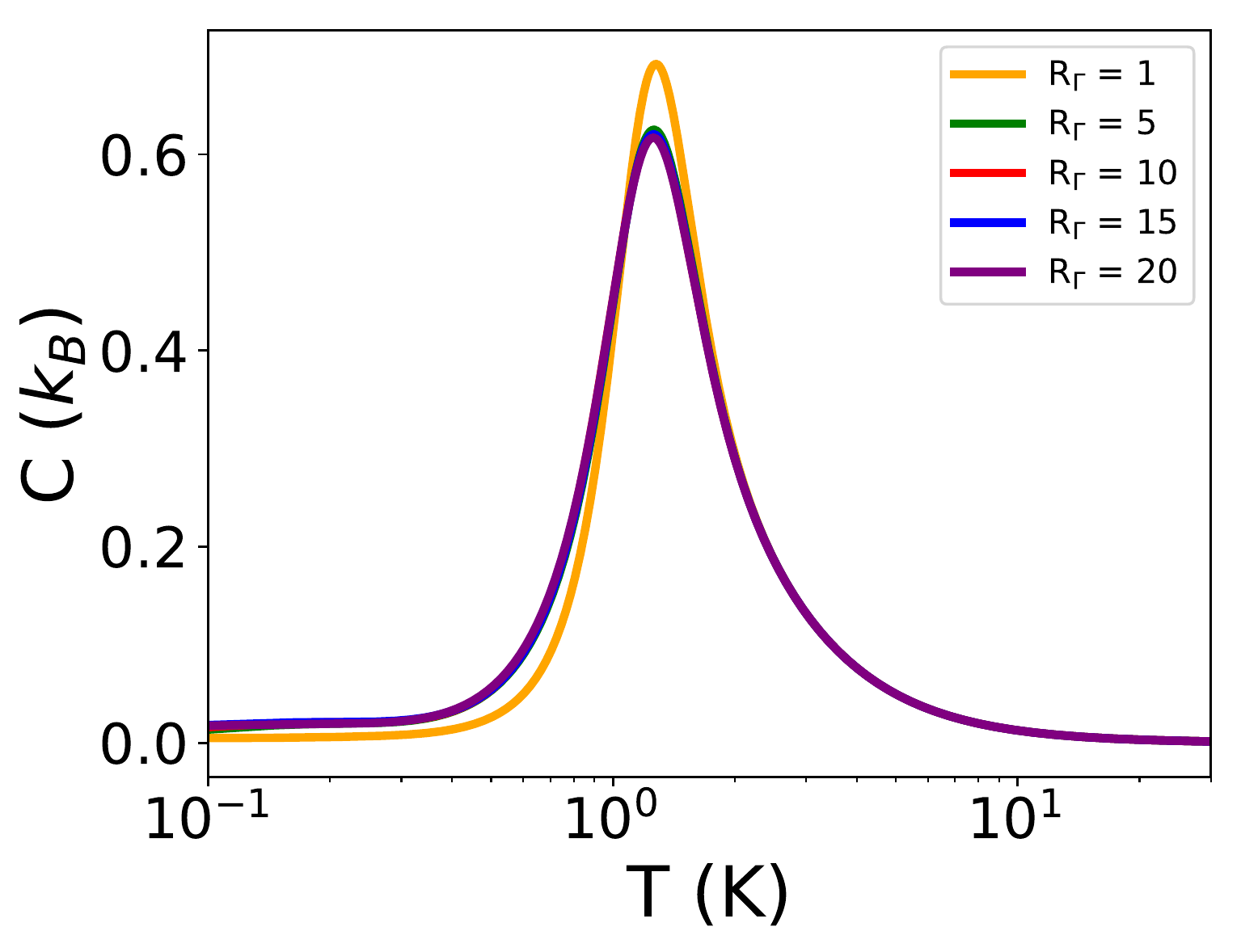}
\caption{(Color online): Convergence of the heat capacity profiles (per mole of the magnetic ion) 
on the $32$ site pyrochlore cluster as a function of temperature. The 
left panel shows results for YbTO for increasing $M$ for fixed $R_{\Gamma}=15$ ($R=120$ total seeds). 
The center and right panels show results for increasing $R_{\Gamma}$ (for fixed $M=100$) for YbTO and ErTO 
respectively.}
\label{fig:C_convergence} 
\end{figure*}	

\section{Details of the numerical calculations}
The $32$ site cluster studied in the paper, has $2$ cells in each direction of the FCC primitive lattice vectors. 
The view of this cluster along the global [111] direction, and the directions in which 
the periodic boundary conditions are applied have been shown in Fig.~1 of the main text. 
We have employed its translational symmetries with momentum directions labelled
${\bf{k}}=(k_1,k_2,k_3)$, the former two representing translations perpendicular to the global [111] 
direction and the latter along [111]. For the $32$ site cluster each $k_i=0$ or $k_i=\pi$ giving rise
to a total of $8$ momentum sectors. 
The Hilbert space in each momentum sector is approximately $536$ million dimensional. 
In Fig.~3 and Fig.~5 of the main text, the lowest lying energies are mapped out in all 
$8$ momentum sectors as a function of parameters in the Hamiltonian (field for Fig.~3 
and $\lambda$ for Fig.~5).

Calculations with each starting random vector took roughly $10$ hours on $48$ cores on the MARCC supercomputer 
for $M=100$ Lanczos iterations. Larger number of iterations were needed ($M=200-400$) 
for studying low energy spectra. Observables that commute with the Hamiltonian (such as the specific heat) 
are calculated using the formulae~\cite{Bronca_review}, 
\begin{eqnarray}
	\langle A \rangle &=&  \frac{1}{Z} \sum_{\Gamma} \frac{N_{\Gamma}}{R_{\Gamma}} \sum_{r=1}^{R_{\Gamma}} \sum_{j=1}^{M} \exp(-\beta E_{jr}) | \langle r | \psi_{jr} \rangle |^2 A_{jr} \\
		Z         &=&  \sum_{\Gamma} \frac{N_{\Gamma}}{R_{\Gamma}} \sum_{r=1}^{R_{\Gamma}} \sum_{j=1}^{M} \exp(-\beta E_{jr}) | \langle r | \psi_{jr} \rangle |^2 
\end{eqnarray}
where $\beta$ is the inverse temperature, $\Gamma$ is a symmetry (sector) index, $N_{\Gamma}$ 
is the (sector) Hilbert space size, $|r \rangle$ is a random vector (in the given sector) 
used to start the Lanczos iteration, $R_{\Gamma}$ is the number of such starting vectors, 
$\psi_{jr}$ is the $j^{th}$ eigenvector obtained after $M$ iterations (with $|r\rangle$ as the start vector), 
$E_{jr}$ is the corresponding Ritz eigenenergy and $A_{jr}=\langle \psi_{jr}|A|r \rangle/\langle \psi_{jr}|r \rangle$

In the main text, we also mentioned that the finite temperature Lanczos method (FTLM) works well because only a 
small number of Krylov space vectors ($M$) and starting vectors ($R_{\Gamma}$ in every symmetry sector) 
are needed to obtain accurate results. (Note that since there are $8$ symmetry sectors, $R=8R_{\Gamma}$). 
To validate this claim we show the convergence properties by varying $M$ and fixed $R_{\Gamma}$, and varying $R_{\Gamma}$ at fixed $M$ 
in Fig.~\ref{fig:C_convergence}. 

For example, Fig.~\ref{fig:C_convergence} shows our results for YbTO when 
we fix $R_{\Gamma}=15$ and vary $M$. As expected, the high temperature features converged the fastest with increasing $M$, 
for example, the entire Schottky anomaly centered at $2.4$ K converges by $M=20$. 
By $M=50$ other lower temperature features have also converged, this is verified by going all the way to $M=100$.
 
In the central panel, we fix $M=100$ and vary $R_{\Gamma}$. It is remarkable that a single random 
vector per sector $R_{\Gamma}=1$ is sufficient for reasonably representing the Schottky anomaly. However, 
to obtain other features quantitatively, one needs $R_{\Gamma} \geq 5$ to converge features 
on the log scale shown. Finer features associated with the "peak" at $T \approx 0.34 $ K show small variations ($0.02$ K) 
between $R_{\Gamma}=5$ to $R_{\Gamma}=15$ but converge by $R_{\Gamma}=20$. 
The right panel of the figure shows analogous results for ErTO. 

\section{Comparison of magnetization profiles}
In the main text, we showed representative magnetization ($M$) profiles as a function of temperature ($T$) 
as part of Fig. 4. Here we clarify that quantum effects are crucial for explaining the trends seen 
in recent experiments in a [111] magnetic field~\cite{Scheie2017}. 

\begin{figure*}
\centering
\includegraphics[width=0.4\linewidth]{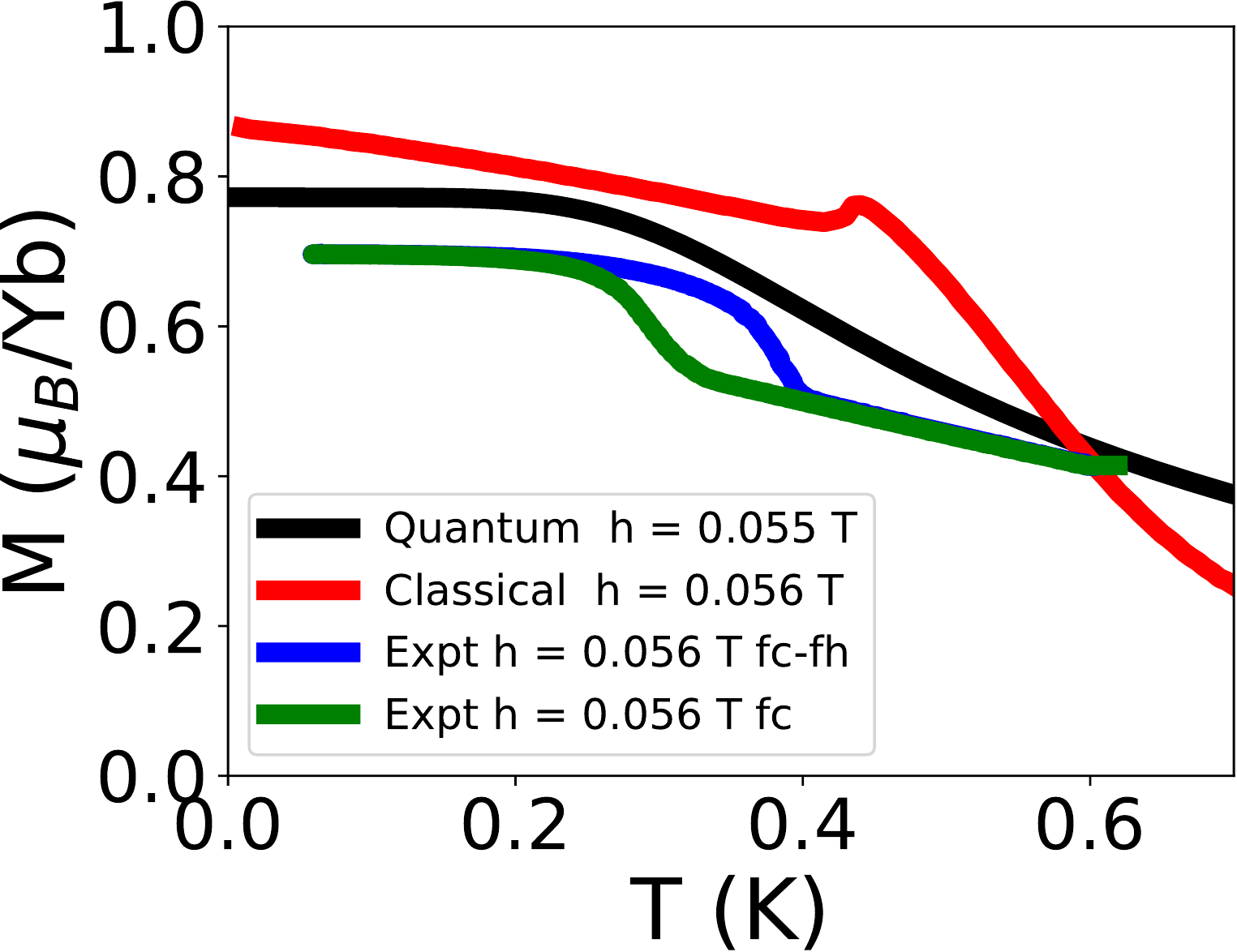}
\includegraphics[width=0.4\linewidth]{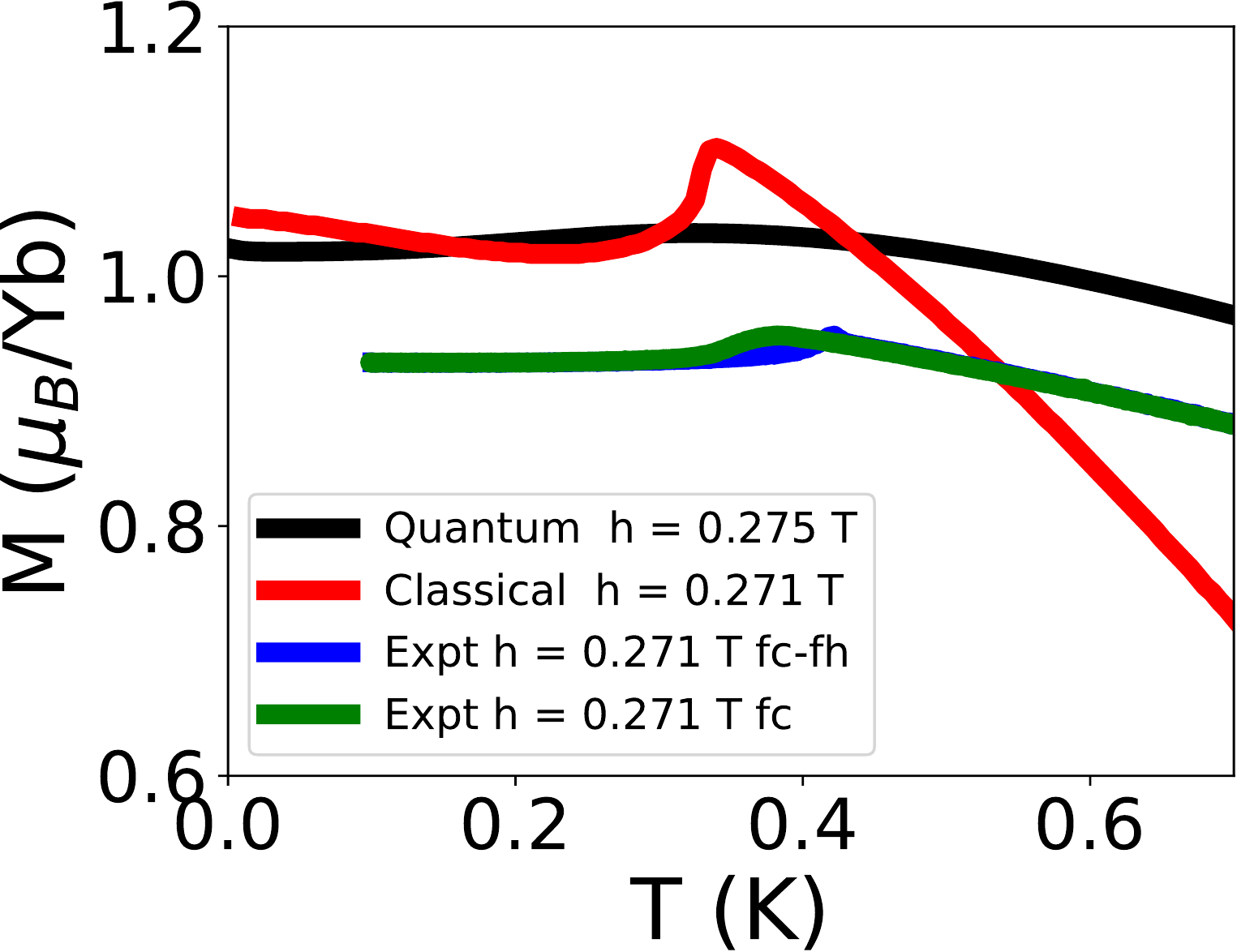}
\caption{(Color online): Quantum and classical magnetization ($M$), 
measured along the same direction as the applied [111] magnetic field ($h$) versus 
temperature ($T$). Comparisons are made to two experimental profiles (with similar field values), left 
for $h=0.055$ T and right for $h=0.275$ T. fc and fh refer to field cooled and field heated measurements respectively.} 
\label{fig:MvsT_supplementary} 
\end{figure*}	

Fig.~\ref{fig:MvsT_supplementary} shows our results for two field strengths ($h$). 
At low temperatures, $M$ vs $T$ is relatively flat, a feature captured quantum mechanically but not classically. 
Moreover, at $h=0.055$ T the classical kink in the magnetization is at much higher temperature, 
consistent with a larger $T_c$ seen classically. In addition, classically, 
the change in $M$ with $T$ is more rapid in the paramagnetic regime in comparison to experiments.
 
In contrast, the quantum calculations largely agree with experiments and correct both these discrepancies. The 
mild disagreements with experiments are attributed to a combination of finite size effects, 
inaccurate parameters and presence of magnetic domains in real systems.

\end{document}